\documentclass[letterpaper]{article} 
\usepackage{aaai2026}  
\usepackage{times}  
\usepackage{helvet}  
\usepackage{courier}  
\usepackage[hyphens]{url}  
\usepackage{graphicx} 
\usepackage{natbib}  
\usepackage{caption} 
\usepackage{caption} 
\usepackage{subcaption} 
\usepackage{algorithm}
\usepackage{algorithmic}
\input{glyphtounicode}
\usepackage{booktabs} 
\usepackage{enumitem}
\usepackage{amsmath}
\usepackage{amsthm}
\usepackage{amsfonts}
\usepackage{dsfont}
\usepackage{cleveref}
\usepackage{multirow}
\usepackage{marginnote}

\graphicspath{{media/}}     

\usepackage{t1enc}
\usepackage{listings}
\lstdefinestyle{promptstyle}{
  basicstyle=\footnotesize\ttfamily,
  breaklines=true,
  columns=fullflexible,   
  keepspaces=true,
  upquote=true,           
  frame=single,
  escapeinside=||,        
  literate=
    {–}{{--}}1            
    {—}{{---}}1           
    {…}{{\ldots}}1
    {’}{{'}}1
    {“}{{``}}1
    {”}{{''}}1
}

\usepackage{xcolor}

\newcommand{\x}{$\mathbb{X}$}

\title{
Auditing Algorithmic Personalization in TikTok Comment Sections
}

\author{
Yueru Yan,
Siqi Wu\\
}

\affiliations {
Indiana University Bloomington\\
\{yueryan, swu2\}@iu.edu
}

\begin{document}

\maketitle

\begin{abstract}
Personalization algorithms are ubiquitous in modern social computing systems, yet their effects on comment sections remain underexplored. In this work, we conducted an algorithmic auditing experiment to examine comment personalization on TikTok. We trained sock-puppet accounts to exhibit left-leaning or right-leaning preferences and successfully validated 17 of them by analyzing the videos recommended on their For You Pages. We then scraped the comment sections shown to these trained partisan accounts, along with five cold-start accounts, across 65 politically neutral videos related to the 2024 U.S. presidential election that contain abundant discussions from both left-leaning and right-leaning perspectives. We find that while the composition of top comments remains largely consistent for all videos, ranking divergence between accounts from different political groups is significantly greater than that observed within the same group for some videos. This effect is strongly correlated with video-level metrics such as comment volume, engagement inequality, and partisan skew in the comment sections. Furthermore, through an exploratory case study, we find preliminary evidence that personalization can result in comment exposure aligned with an account's political leaning. However, this pattern is not universal, suggesting that the extent of politically oriented comment personalization is context-dependent.
\end{abstract}

\begin{links}
    \link{Code}{https://github.com/raye22/tiktok_personalization}
\end{links}

\section{Introduction}

TikTok has emerged as a primary source of information, with one-fifth of U.S. adults now regularly consuming news on the platform~\cite{tomasik_matsa_2025_tiktok_news}. At the core of TikTok's success is its recommendation system, which delivers an endless stream of personalized content for maximizing user engagement~\cite{zannettou2024analyzing}. Yet the pervasive reliance on opaque algorithms has raised pressing concerns about public trust, information fairness, and platform accountability~\cite{shin2019role}. Recently, a growing body of research has leveraged a technique called ``algorithmic auditing'' to scrutinize recommendations across social media platforms, with a particular focus on ``For You Page'' (FYP) feeds~\cite{boeker2022empirical,srba2023auditing,mousavi2024auditing} and search results~\cite{juneja2023assessing,jung2025algorithmic}. These studies have identified significant issues, such as potential rabbit holes on mainstream topics~\cite{gong2025clipmind} and a slant toward Republican content during the 2024 U.S. presidential election~\cite{ibrahim2025tiktok}. 

However, platform recommendation extends beyond FYP feeds and search interfaces to comment sections, where ranking mechanisms have evolved from simple chronological or popularity-based sorting to complex, algorithmically mediated systems. For example, Reddit highlights ``Best'' comments, YouTube promotes ``Top'' comments, and Facebook prioritizes ``Most relevant'' comments, all of which are curated by opaque, proprietary algorithms rather than transparent rules. Industry insights suggest that this shift is strategic, as platforms like Kuaishou, a short-video platform similar to TikTok, have begun optimizing for ``comment staytime'' and applying personalized ranking to comment sections to increase user engagement~\cite{zhang2025comment}.

Since public comment sections serve as central spaces for people to share opinions and engage in conversations~\cite{diakopoulos2011towards,park2016supporting}, personalized comment exposure can create a filter bubble by selectively elevating viewpoints that align with a user's existing beliefs~\cite{flaxman2016filter}. This risk is particularly acute on TikTok. Unlike platforms that allow users to toggle between alternative options such as chronological or popularity-based ordering, TikTok only displays a single, algorithmically determined order~\cite{TikTokSupport}. This design forces users into a largely passive consumption role and limits their ability to opt out of algorithmic curation.

While prior auditing research has extensively examined recommendation systems and search engines, little is known about personalization in comment sections. To address this gap, we present, to our knowledge, the first auditing study of comment personalization effects on TikTok. We pose two research questions. \textbf{RQ1}: Does algorithmic personalization exist in TikTok comment sections? \textbf{RQ2}: In the context of political discussions, does personalization favor comments that align with a user's political leaning?

To answer them, we developed a ``train-validate-evaluate'' framework. First, we instantiated 20 sock-puppet accounts using a ``search-then-watch'' strategy targeting prominent partisan accounts (e.g., MSNBC, Fox News) to mimic real world users with certain political leanings. We also created five cold-start accounts as a control group. Second, we validated the platform-perceived political leaning of these accounts by manually examining the political leaning of videos recommended on their FYPs. 17 accounts (nine left-leaning and eight right-leaning) successfully met the validation criteria. Third, we scraped the on-screen comments and their displayed order from a set of politically neutral videos with abundant and diverse partisan discussion. We manually annotated each comment's stance to characterize the partisan composition of the ranked lists. To quantify personalization effects, we computed two metrics: Jaccard Distance (JD) to measure overlap in the sets of visible comments, and Normalized Damerau–Levenshtein Distance (NDLD) to capture dissimilarities in comment ordering. We applied the Analysis of Similarities (ANOSIM) test to quantify the degree of personalization as the difference between within-group and between-group variation. This design enables us to examine whether politically dissimilar users are systematically exposed to different comments on the same videos.

Our findings reveal a nuanced picture of comment personalization on TikTok. We find that personalization is statistically significant in the ranking of comments, but not in the composition of comment sections across different political groups. While the set of visible comments remains largely consistent across audit groups (mean JD $< 0.08$), the order in which they appear varies significantly, with 25\% of videos showing high ranking divergence. However, this effect is not universal; rather, it is highly conditional and strongly associated with specific video characteristics. Spearman's rank correlation analysis indicates that ranking divergence between opposing political groups is primarily associated with the imbalance of partisan discussion, whereas divergence relative to the control group increases with comment volume and engagement inequality. Regarding political alignment, our exploratory case study on five videos with diverse political discussions suggests that personalization can alter the visibility of certain partisan viewpoints. 

The main contributions of this work are as follows:
\begin{itemize}
    \item We present, to our knowledge, the first investigation of algorithmic personalization in TikTok comment sections.
    \item We show that, although users from different political groups are exposed to largely the same set of comments, differences in comment ranking can produce divergent exposures. However, this effect is conditional on specific videos rather than a universal platform phenomenon.
    \item We develop a methodological framework based on the Analysis of Similarities test to compare within-group and between-group similarity. This approach, rarely used in algorithmic auditing research, offers a robust tool for analyzing ranked lists in personalized systems.
\end{itemize}


\section{Related Work}

Online platforms increasingly deploy recommendation algorithms across a wide range of contexts. To understand how these algorithms shape the information users encounter, researchers have turned to auditing studies. Broadly, this literature falls into two categories. The first focuses on search interfaces, such as Google Search~\cite{le2019measuring,jung2025algorithmic} and YouTube Search~\cite{hussein2020measuring,juneja2023assessing}, examining whether identical queries yield different results for users with distinct personal attributes (e.g., search history, location, or other profile characteristics)~\cite{krafft2019did}. The second focuses on user homepage feeds~\cite{bandy2021more,chen2023subscriptions,ye2025auditing}. Previous work has investigated platforms such as YouTube, TikTok, \x, and news applications to understand how algorithms curate personalized feeds and whether such curation contributes to information filter bubbles~\cite{tomlein2021audit,vombatkere2024tiktok,ibrahim2025tiktok}. Auditing has become a widely used approach due to the limited availability of real user data. It involves creating controlled user profiles that simulate different characteristics or behaviors and comparing the system outputs between them~\cite{asplund2020auditing,boeker2022empirical,liu2024train}.

In political contexts, auditing studies have produced substantial, albeit sometimes mixed, evidence of algorithmic personalization across platforms. \citet{le2019measuring} found that personalization reinforced the presumed partisanship of users, with Google News results skewing left for liberal-trained profiles and right for conservative-trained profiles, echoing concerns about filter bubbles. Similarly, a six-week audit of \x's homepage timeline found that both left- and right-leaning users saw predominantly in-party content and experienced reduced exposure to opposing viewpoints~\cite{ye2025auditing}. In contrast, \citet{juneja2023assessing} found no evidence of personalization in YouTube search following the enactment of regulation policies regarding election misinformation.

TikTok has likewise received increasing scrutiny as researchers seek to understand the mechanisms driving personalization in short-video feeds~\cite{vombatkere2024tiktok,mousavi2024auditing,gong2025clipmind}. \citet{boeker2022empirical} found that the ``follow'' behavior was the strongest driver of personalization, followed by ``likes'' and ``watch time''. This finding is consistent with \citet{zannettou2024analyzing}, who analyzed data donations from 347 TikTok users and showed that engagement behaviors quickly narrowed content exposure, with ``follows'' and ``likes'' leading to more homogeneous FYPs. More recent work suggests that these effects vary across content domains. \citet{gong2025clipmind} introduced the ``ClipMind'' framework and found that, for broad interest categories such as Food or Beauty, TikTok's algorithm increasingly recommended similar content, indicating strong reinforcement of user preferences. In contrast, for more niche topics such as War or Mental Health, they found no evidence of algorithmic rabbit holes that may lead to extreme content.

\begin{figure*}[tb]
    \centering
    \includegraphics[width=1\linewidth]{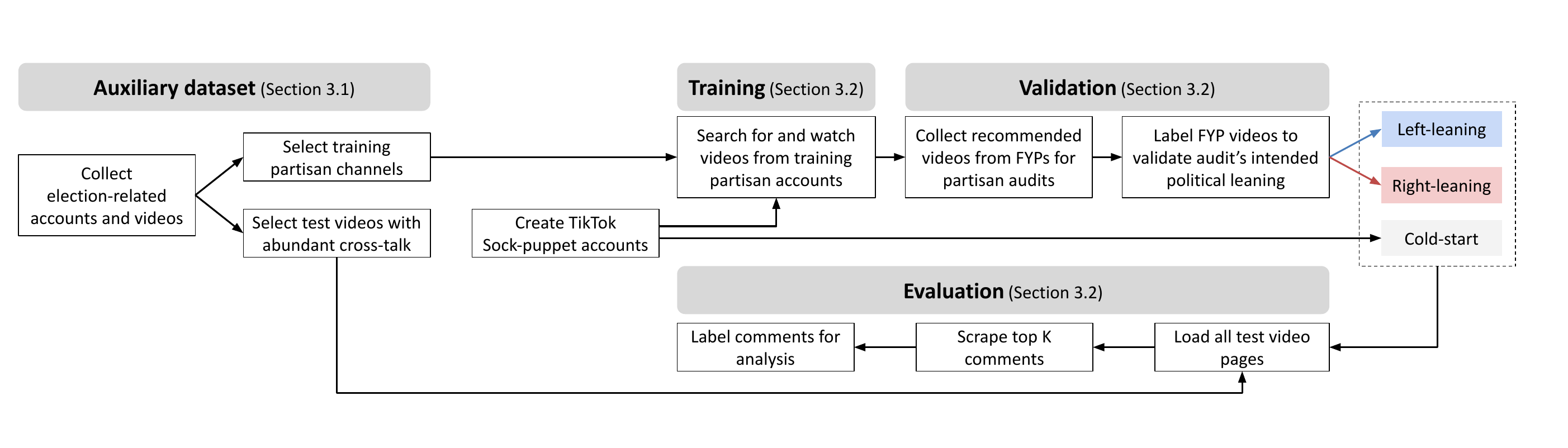}
    \caption{Overview of the experiment workflow. The process begins with constructing an auxiliary dataset to identify valid partisan channels for training and center videos for testing. We then follow a ``train-validate-evaluate'' framework: (1) training sock-puppet accounts to perform ``search-then-watch'' actions for selected partisan channels, (2) validating their platform-perceived political leanings by examining the videos recommended on FYP feeds, and (3) scraping top-level comments and their ordering from test videos to measure how the algorithm alters comment visibility for different political groups.}
    \label{fig:audit_design}
\end{figure*}

Despite this progress, comment sections remain largely underexplored in the auditing literature. To date, research on comment sections has focused primarily on the content and dynamics of user discussions~\cite{diakopoulos2011towards,anderson2014nasty, alafwan2023comments}, rather than on the algorithms that curate comment visibility and ordering. We address this gap by shifting attention to algorithmic personalization and by conducting, to our knowledge, the first sock-puppet auditing study of TikTok comment sections in the context of the 2024 U.S. presidential election. Although prior work has identified follow and like behaviors as strong drivers of personalization~\cite{boeker2022empirical}, we deliberately adopted a ``search-then-watch'' strategy. This design choice was motivated by both ethical and methodological considerations. Ethically, it minimizes unintended platform manipulation by avoiding the artificial inflation of follower or like counts that could affect real users. Methodologically, our setting more closely resembles search auditing than feed auditing: unlike passive feed consumption, our sock-puppet accounts must proactively open a video page and then navigate to its comment section, rather than passively receive recommendations through a continuous feed.

\section{Methodology}
\label{sec:methodology}

To investigate personalization in TikTok comment sections, we designed a controlled sock-puppet auditing experiment. Following~\citet{sandvig2014auditing}, sock-puppet audits are automated computer programs that mimic user behavior to interact with the system while simultaneously recording system responses. This technique is widely used to examine black box components (e.g., recommendation algorithms) in social computing systems~\cite{bandy2021more,juneja2023assessing}. \Cref{fig:audit_design} illustrates the overall workflow, which consists of three phases: training, validation, and evaluation. Because both the training and evaluation phases required specific TikTok content related to the 2024 U.S. presidential election, we first constructed an auxiliary dataset. The auditing experiment was conducted on April 15, 2025, and each phase was initiated immediately after the preceding phase was completed.

\subsection{Dataset Construction}
\label{ssec:method_dataset}

\subsubsection{TikTok Political Channels and Videos Dataset} 
To construct our TikTok political dataset, we derived a list of target accounts from a YouTube political discussion dataset~\cite{wu2021cross}, which includes 1,278 U.S. media outlets classified as left-leaning, right-leaning, or center. For clarity, we refer to these media outlets as ``channels'' and to our sock-puppet audits as ``accounts'' throughout this paper.

To identify the corresponding TikTok channel IDs, we conducted Google searches using queries of the form ``\{YouTube channel name\} TikTok account''. From the top five search results, we extracted the links matching the format ``https://www.tiktok.com/$@{TikTok\_ID}$'' and obtained the associated TikTok ID. This process yielded an initial set of 1,187 TikTok channels. To ensure matching accuracy, we used Python's \texttt{SequenceMatcher} to compute the text similarity between each YouTube channel name and its corresponding TikTok profile name. For pairs with a similarity score below 0.5, we manually verified them by checking whether the YouTube channel description linked to the TikTok profile. Pairs without such links were excluded, resulting in 971 TikTok channels. 

To retain only election-related videos posted by these 971 channels, we constructed a keyword list through a two-step process. First, we defined a set of seed hashtags based on major political candidates, including ``harris'', ``kamala'', ``trump'', ``kamalaharris'', and ``donaldtrump''. Second, we collected videos containing these tags between 2024-09-09 and 2024-09-11, coinciding with the second U.S. presidential debate (September 10, 2024), a period expected to generate a high volume of election-related content. From this dataset, we identified the most frequently used hashtags. Among the top 20, we retained election-related terms such as ``vote2024'', ``democrat'', and ``trump2024'', while excluding generic or unrelated tags like ``usa'' and ``fyp''. This process yielded a final list of 17 election-related hashtags.

Finally, we extracted election-related videos from the 971 matched TikTok channels during the 30-day period preceding the 2024 election (October 7 to November 5, 2024), using the hashtag list constructed in the previous step. We retained 346 channels that published election-related content during this period. To ensure the highest accuracy, we conducted a comprehensive manual validation. This left us with 257 verified TikTok channels and 4,691 election-related videos. The final dataset contains 3,244 videos from 166 left-leaning channels, 825 videos from 50 right-leaning channels, and 622 videos from 41 center channels. 

\subsubsection{Selecting Test Videos} 
In this study, our objective is to examine whether a user's political leaning affects the comments shown to them. However, if a comment section is dominated by a single partisan perspective, personalization effects may be difficult to detect due to the limited availability of diverse viewpoints. To address this, we identify a set of videos with non-trivial cross-partisan discussions. 

We focused on videos posted by center TikTok channels, which prior work suggests attract more heterogeneous audiences~\cite{chae2024cross}. This selection included 622 videos posted by the 41 center channels identified in the previous step. To further assess the composition of these videos' comment sections, we used the TikTok API to collect all top-level comments for each video. Since TikTok's web interface displays only five to six comments per screen, we retained only videos with at least 50 comments to ensure sufficient data for analysis, resulting in 239 videos.

Because manual labeling was infeasible for the entire dataset, we needed a computational approach to estimate the fraction of partisan comments on each center video. Two authors first collaboratively reviewed a random subset of comments. From this process, we made a few observations for labeling the political stance of TikTok comments: (a) TikTok comments are typically short and often contain pronouns; (b) pronouns can only be parsed by examining the video content; (c) a user's stance (support vs. oppose) is usually discernible from the comment itself. Based on these observations, we developed an annotation codebook with five labels: Pro-Democrat, Anti-Democrat, Pro-Republican, Anti-Republican, and Neutral. The codebook is detailed in Appendix~\ref{app:comment_codebook}. These five labels were later collapsed into three broad categories: Left-leaning (Pro-Democrat and Anti-Republican), Right-leaning (Pro-Republican and Anti-Democrat), and Neutral. We then used a large language model (LLM), GPT-4o, to classify the political stance of each comment using the prompt described in Appendix~\ref{app:prompt}. To account for the context-dependency of pronouns, we provided the model with both the comment text and the corresponding video description. 

For each of the 239 videos, we randomly sampled 50 comments and classified them using GPT-4o. This procedure provides an unbiased estimate of the fraction of partisan comments on each video. We included a video in the test set if at least 25\% of the sampled comments were classified as left-leaning and at least 25\% as right-leaning. Applying this criterion yielded 67 test videos from nine unique channels.

\subsubsection{Classifier Accuracy Validation}
To assess the reliability of our automated labeling, we evaluated the performance of GPT-4o against a human-annotated gold standard. Two authors independently labeled a shared sample of 100 comments using the codebook, achieving a Cohen's $\kappa$ of 0.78. Disagreements were resolved through discussion to produce a gold standard labeled dataset. We then benchmarked GPT-4o against this dataset. The model achieved an accuracy of 0.84 and a Cohen's $\kappa$ of 0.79, indicating strong agreement with human annotators. These metrics exceed commonly accepted thresholds for inter-rater reliability, supporting the use of GPT-4o for large-scale classification of political comments in the evaluation phase.

\subsection{Sock-Puppet Audit Design}
\label{ssec:method_training}

In this study, we created 25 TikTok sock-puppet accounts to simulate users aged 22 to 25, reflecting the platform's core young adult demographic. Among these, ten accounts were behaviorally trained to exhibit left-leaning preferences, ten to exhibit right-leaning preferences, and five served as a cold-start control group with no prior interaction history.

\begin{table}[t]
\small
\centering
\begin{tabular}{l r r}
\toprule
\multirow{2}{*}{\textbf{Channel type}} & \textbf{Left-leaning ID} & \textbf{Right-leaning ID} \\
 & \textbf{(\#followers)} & \textbf{(\#followers)} \\
\midrule
\multirow{2}{*}{Election candidate} & kamalaharris & realdonaldtrump \\
 & (9,300,000) & (14,600,000) \\
\multirow{2}{*}{Cable news} & cnn & foxnews \\
 & (6,400,000) & (1,700,000) \\
\multirow{2}{*}{Cable news} & msnbc & newsmaxtv \\
 & (3,900,000) & (162,100) \\
\multirow{2}{*}{Independent media} & democracynow.org & dailywire \\
 & (2,971,000) & (3,600,000) \\
\multirow{2}{*}{Political influencer} & davidpakmanshow & real.benshapiro \\
 & (855,800) & (2,700,000) \\
\bottomrule
\end{tabular}
\caption{List of paired TikTok channels used to train sock-puppet accounts. To ensure balanced partisan signals, five left-leaning and five right-leaning channels were matched by media type and follower count.}
\label{tab:tiktok_accounts_by_type}
\end{table}

\subsubsection{Training Phase}
In a pilot experiment, we randomly sampled a large number of left-leaning (similarly, right-leaning) election-related videos and made sock-puppet accounts passively watch them. However, the validation results indicated that passive consumption of partisan content alone was insufficient to induce meaningful political personalization. We therefore turned to an active ``search-then-watch'' strategy, hypothesizing that explicitly searching for known partisan channels would provide a stronger signal of user intent.

To implement this strategy, we first identified prominent partisan channels as search targets. From the pool of 257 TikTok channels, we ranked them by follower count and examined the top 20 channels on both left- and right- leaning side. We then manually selected channels by type (e.g., candidate, cable news, or independent creators) and paired left-leaning channels with right-leaning counterparts of the same type and comparable follower counts. This matching strategy ensures that, while leveraging high profile channels to facilitate active search, the popularity signal remains comparable across audit groups, leaving political leaning as the primary distinguishing factor. This process produced ten partisan channels, five per side. As shown in~\Cref{tab:tiktok_accounts_by_type}, each side includes one election candidate, two partisan cable news outlets, one independent media, and one political influencer.

The training procedure was executed as follows. First, each sock-puppet account was preassigned a target political leaning. The account then logged into TikTok's web interface via Selenium, queried the search bar with the name of a selected channel matching the assigned leaning, and navigated to its profile page. It then selected the first video displayed (i.e., either the most recent or a pinned video) and advanced through subsequent videos by pressing the ``Down'' arrow key. After watching ten videos from a given channel, the account returned to the search interface, proceeded to the next target channel, and repeated the ``search-then-watch'' actions. This process continued until all five target partisan channels had been covered. In total, each partisan account performed five searches and watched 50 videos per training day, for a duration of three days.

\subsubsection{Validation Phase} 
We validated whether TikTok had indeed learned the intended political leaning of each trained sock-puppet account by examining their FYP recommendations. For example, left-leaning accounts are expected to be recommended more left-leaning videos than right-leaning ones. Because TikTok videos auto-play upon loading, and such interactions may inadvertently influence the recommendation algorithm, we designed a procedure to collect recommended videos without triggering ``watch'' events. Specifically, we sent web requests to TikTok's servers from logged-in accounts, parsed the first ten recommended video IDs from the network responses, and immediately terminated the session. This process was repeated ten times per account. On average, we obtained approximately 60 recommended videos per sock-puppet account after training, as FYP requests occasionally failed to return results.

Next, we analyzed the political leaning of the recommended videos. Using the collected video IDs, two authors manually accessed the corresponding URLs via a separate web browser to review the content. To support consistent annotation, the authors first independently reviewed a subset of videos and then collaboratively developed an annotation codebook. The codebook defines four categories: Left-leaning, Right-leaning, Neutral, and Non-political. We defined left-leaning and right-leaning videos as those explicitly reflecting partisan positions. For example, immigration-related content supporting immigrant rights was labeled left-leaning, whereas religion-related content promoting Christianity was labeled right-leaning. Neutral videos referenced political topics but did not express a clear stance, while non-political videos were unrelated to politics (e.g., entertainment, lifestyle, or comedy). Based on a random sample of 194 videos, the two annotators achieved a Cohen's $\kappa$ of 0.91, indicating near-perfect agreement. Disagreements were subsequently resolved through discussion, and the finalized codebook was used to label the remaining videos, with each author independently annotating half of the dataset. For each account, we computed the proportion of recommended videos in each category. Prior work suggests that TikTok's FYP typically allocates 30-50\% of content to reinforcing known user interests, with the remainder introducing novel content~\cite{vombatkere2024tiktok}. Based on this observation, we established the following validation criteria: an account was considered successfully trained if more than 35\% of its FYP content was political and the proportion of videos matching the target leaning exceeded that of the opposing leaning. If an account did not meet this criterion, we extended its training by another day and repeated the validation process, for up to two additional days. Accounts that still failed to meet the criteria after retraining were excluded.

Our training protocol proved highly effective, successfully validating 17 of the 20 initial accounts: nine left-leaning (hereafter $L$) and eight right-leaning (hereafter $R$). Among these, three $L$ accounts required one additional day of training, and one $R$ account required two additional days to meet the validation criteria. On average, 45\% of the videos recommended to the nine $L$ accounts were left-leaning, compared to only 5\% that were right-leaning. For the eight $R$ accounts, 41\% of their FYP videos were right-leaning, while only 6\% were left-leaning. We attribute the effectiveness of this training strategy to the explicit nature of the ``search'' action. Whereas passive watching relies on noisy engagement signals such as dwell time, searching for accounts like ``foxnews'' provides a direct and unambiguous signal of user intent, enabling the algorithm to infer user political preferences more rapidly.

\subsubsection{Evaluation Phase}
We audited the 67 test videos with abundant cross-partisan discussions identified in~\Cref{ssec:method_dataset} using the 17 validated accounts and five cold-start control accounts. For each video, the accounts navigated to the comment section and performed up to 20 scroll down actions. During this process, a script recorded each comment's text, rank position, like count, and reply count for all top-level comments. The session ended either after 20 scroll down actions were completed or when no new comments were loaded. Although we initially targeted 67 videos, technical glitches prevented comment collection for two videos. Consequently, the final analytic dataset consists of the first 50 comments observed by 22 accounts across 65 test videos.

\subsection{Metrics to Measure Personalization}
\label{ssec:method_evaluation}

This section introduces the metrics used to quantify personalization. Following prior work on algorithmic auditing~\cite{le2019measuring,juneja2023assessing}, we computed two distance metrics for every pair of ranked comment lists. These two metrics capture dissimilarity from different perspectives and jointly provide a comprehensive assessment.

\begin{itemize}
    \item Jaccard Distance (JD): It measures the fraction of distinct items between two sets. Formally, for video $v$, let $S_{v}(a)$ and $S_{v}(b)$ denote the sets of comments shown to accounts $a$ and $b$. $\mathrm{JD}_{v}(a,b)$ is defined as $1 - \frac{\left| S_{v}(a) \cap S_{v}(b) \right|}{\left| S_{v}(a) \cup S_{v}(b) \right|}$. JD ranges from 0 to 1, where 0 indicates that the two sets are identical and 1 indicates that they are disjoint.
    
    \item Normalized Damerau--Levenshtein Distance (NDLD): It measures the edit distance between two ranked lists. Formally, for video $v$ and accounts $a$ and $b$, let their top-$k$ ranked comment lists be $L_{v}(a)$ and $L_{v}(b)$. $\mathrm{NDLD}_v(a,b)$ is defined as $\frac{d_{\mathrm{DL}}\big(L_{v}(a), L_{v}(b)\big)}{k}$, where $d_{\mathrm{DL}}$ counts the minimum number of operations (i.e., insertions, deletions, substitutions, and transpositions) required to transform one list into the other. NDLD is between 0 and 1, where 0 indicates that the two ranked lists are identical and 1 indicates maximal dissimilarity (e.g., requiring substituting every single item).
\end{itemize}

However, differences between groups alone are insufficient to conclude personalization, as similar levels of variation may also arise within groups. Therefore, we require a metric that accounts for both between-group and within-group differences. We adopt the Analysis of Similarities (ANOSIM), a non-parametric statistical test that evaluates whether between-group dissimilarities are greater than within-group dissimilarities based on a dissimilarity matrix~\cite{clarke1993non}. Formally, let $N$ be the total number of accounts for a video, $\bar r_B$ and $\bar r_W$ be the mean ranks of between-group and within-group dissimilarities, respectively. The ANOSIM $\mathbb{R}$ statistic is defined as
\[
\mathbb{R} =\frac{\bar r_B-\bar r_W}{\,\tfrac{1}{4}N(N-1)\,}\,
\]

$\mathbb{R}$ is between -1 and 1, and is computed separately for each video. To interpret $\mathbb{R}$ in our context, we use $\mathbb{R}_{L,R}$ as an example, which quantifies the differences in the comment lists between the $L$ and $R$ accounts. A positive value ($\mathbb{R}_{L,R} > 0$) implies evidence of personalization, meaning that the dissimilarity between $L$ and $R$ accounts exceeds the internal variation observed within either group. Conversely, a value near or below zero implies no personalization, as the divergence between political groups is indistinguishable from, or smaller than, the variation observed within them.

We assessed statistical significance by generating a null distribution through 1,000 random permutations of the comment lists. This approach enables us to evaluate whether the observed differences between sock-puppet groups are statistically significant compared with variation within groups.

\section{Analyses and Results}
\label{sec:finding}

In this section, we address our two research questions by focusing on the top-$10$ comments. We selected this threshold to approximate typical user consumption on TikTok and to align with sampling depths used in prior auditing studies~\cite{le2019measuring,jung2025algorithmic}. To ensure robustness, we replicated all analyses using the top-$50$ comments and observed qualitatively similar results.

\subsection{RQ1: Does Comment Personalization Exist?}
\label{ssec: finding_rq1}

\begin{figure}[tb]
  \centering
  \begin{minipage}{0.23\textwidth}
    \centering
    \includegraphics[width=\linewidth]{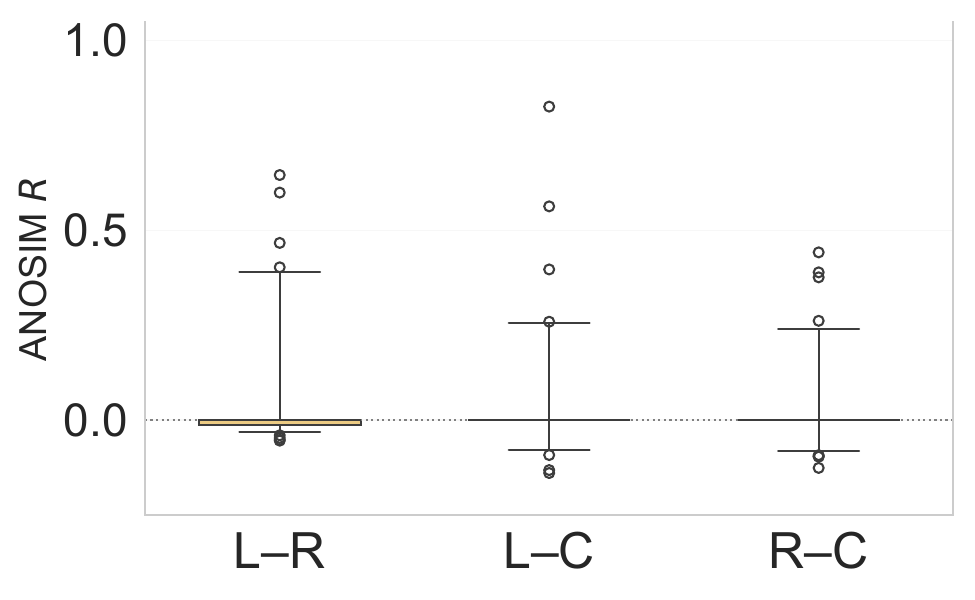}
    \caption*{(a) $\mathbb{R}$(JD)}
    \label{fig:jd_r_comparsion}
  \end{minipage}
  \hfill
  \begin{minipage}{0.23\textwidth}
    \centering
    \includegraphics[width=\linewidth]{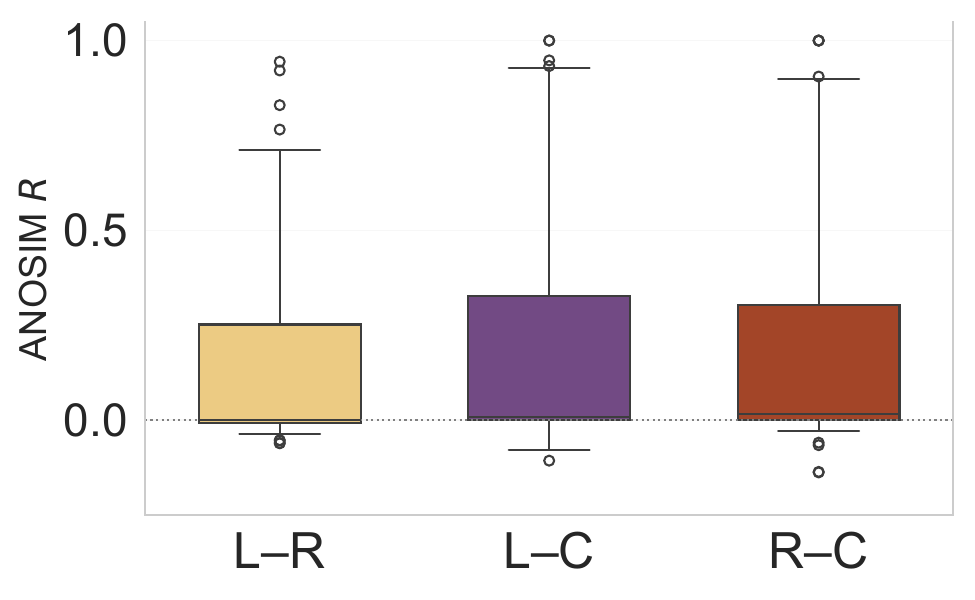}
    \caption*{(b) $\mathbb{R}$(NDLD)}
  \end{minipage}
  \caption{Distribution of personalization measures across all test videos. The plots show ANOSIM $\mathbb{R}$ values for pairwise comparisons between audit groups using (a) JD and (b) NDLD. The low JD values in (a) indicate that all accounts were exposed to nearly identical sets of comments, whereas the greater variance in NDLD in (b) reveals substantial differences in comment ranking for a subset of videos.}
  \label{fig:anosim_results}
\end{figure}

As described in~\Cref{ssec:method_evaluation}, we applied the ANOSIM $\mathbb{R}$ statistic to measure personalization in both comment composition (JD) and ranking order (NDLD).

For each test video, we collected 22 ranked comment lists from three audit groups: nine $L$ accounts, eight $R$ accounts, and five cold-start (hereafter $C$) accounts. This resulted in $(22 \times 21)/2 = 231$ pairwise comparisons per video, including 157 between-group comparisons (e.g., $L$ vs. $R$) and 74 within-group comparisons (e.g., $L$ vs. $L$). We then computed the ANOSIM $\mathbb{R}$ statistic for each video across three pairwise groupings: $\mathbb{R}_{L,R}$, $\mathbb{R}_{L,C}$, and $\mathbb{R}_{R,C}$.

\Cref{fig:anosim_results}(a) presents the results for content composition, with all $\mathbb{R}$(JD) values tightly concentrated around zero. This indicates that differences in comment composition between groups are not greater than the variation within groups. To further validate this finding, we examined the raw JD values for every video and found that the mean JD was consistently below 0.08. Taken together, these results suggest that there is no personalization in the composition of TikTok comment sections, as all accounts were exposed to a nearly identical pool of comments regardless of their political leanings.

The low $\mathbb{R}$(JD) values suggest that any differences captured by $\mathbb{R}$(NDLD) are primarily from variation in comment ranking rather than composition. \Cref{fig:anosim_results}(b) shows the results for ranking divergence. Although the median values remain close to zero, the distribution exhibits a long tail, with more than 25\% of videos having $\mathbb{R}$(NDLD) greater than 0.25. These findings indicate a heterogeneous pattern. For most videos, comment ordering is largely consistent across users. However, for a subset of videos, accounts with different political leanings are exposed to substantially different ranking orders compared to the variation within their own groups. We conclude that personalization exists in comment ranking, but this effect is not universal. This prompts us to investigate the patterns of variation across videos.

\subsection{Variation of Personalization Across Videos}
\label{ssec:video_clustering}

To better understand why personalization emerges in some contexts but not others, we examine the relationship between observed ranking divergence and the structural characteristics of comment sections. Specifically, we compute four features for each video:

\begin{itemize}
    \item Volume of discussion: The total number of comments, log-transformed to reduce skewness.
    \item Inequality of comment replies: The concentration of discussion depth, measured using the Gini coefficient of comment reply counts.
    \item Inequality of comment likes: The concentration of user engagement, measured using the Gini coefficient of the comment like counts.
    \item Imbalance of partisan discussion (IPD): The absolute difference between the normalized proportions of left-leaning and right-leaning comments, based on GPT-4o classifications of 50 randomly sampled comments.
\end{itemize}

\begin{figure}[tb]
  \centering
  \includegraphics[width=0.98\linewidth]{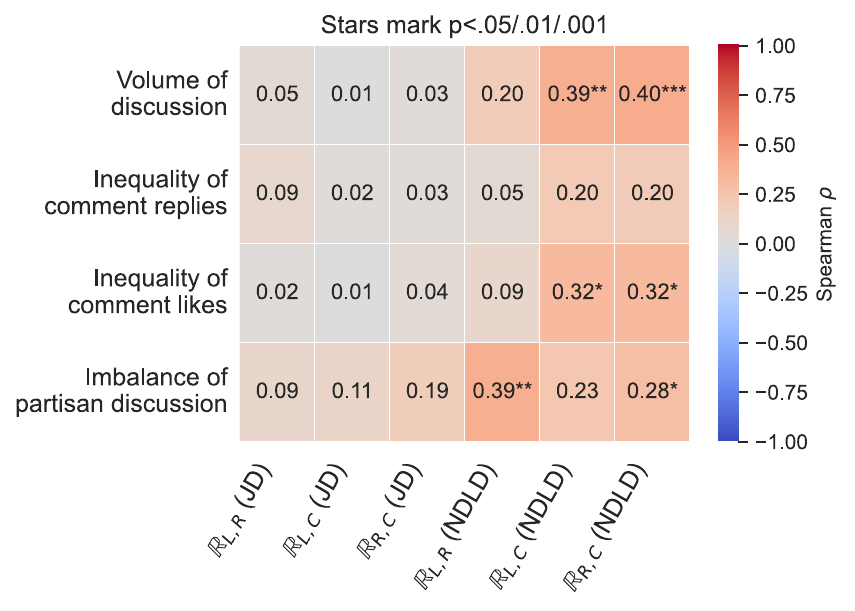}
  \caption{Heatmap of Spearman's rank correlations between video characteristics and personalization metrics. Ranking divergence (NDLD columns) increases significantly with discussion volume, engagement inequality, and imbalance of partisan discussion, whereas content composition (JD columns) shows little to no correlation with these factors.}
  \label{fig:corr_analysis}
\end{figure}

\subsubsection{Correlation Analysis}
\Cref{fig:corr_analysis} presents the Spearman's rank correlations between these features and the ANOSIM $\mathbb{R}$ personalization metrics. Three key patterns emerge. First, $\mathbb{R}$(JD) shows no significant correlation with any features, reinforcing our finding from RQ1 that comment composition remains stable regardless of the underlying discussion dynamics. Second, for ranking divergence, $\mathbb{R}_{L,R}$(NDLD) is positively correlated only with the imbalance of partisan discussion, suggesting that $L$ and $R$ accounts are exposed to more divergent comment rankings on videos with politically skewed discussions. Third, $\mathbb{R}_{L,C}$(NDLD) and $\mathbb{R}_{R,C}$(NDLD) exhibit a similar pattern: as comment volume and inequality of likes increase, both $L$ and $R$ accounts are exposed to more divergent rankings compared to cold-start accounts. In contrast, the inequality of comment replies shows no correlation with any personalization metrics.

\subsubsection{Video Clustering}
\begin{figure}[tb]
  \centering
  \begin{minipage}[t]{0.22\textwidth}
    \includegraphics[width=\textwidth]{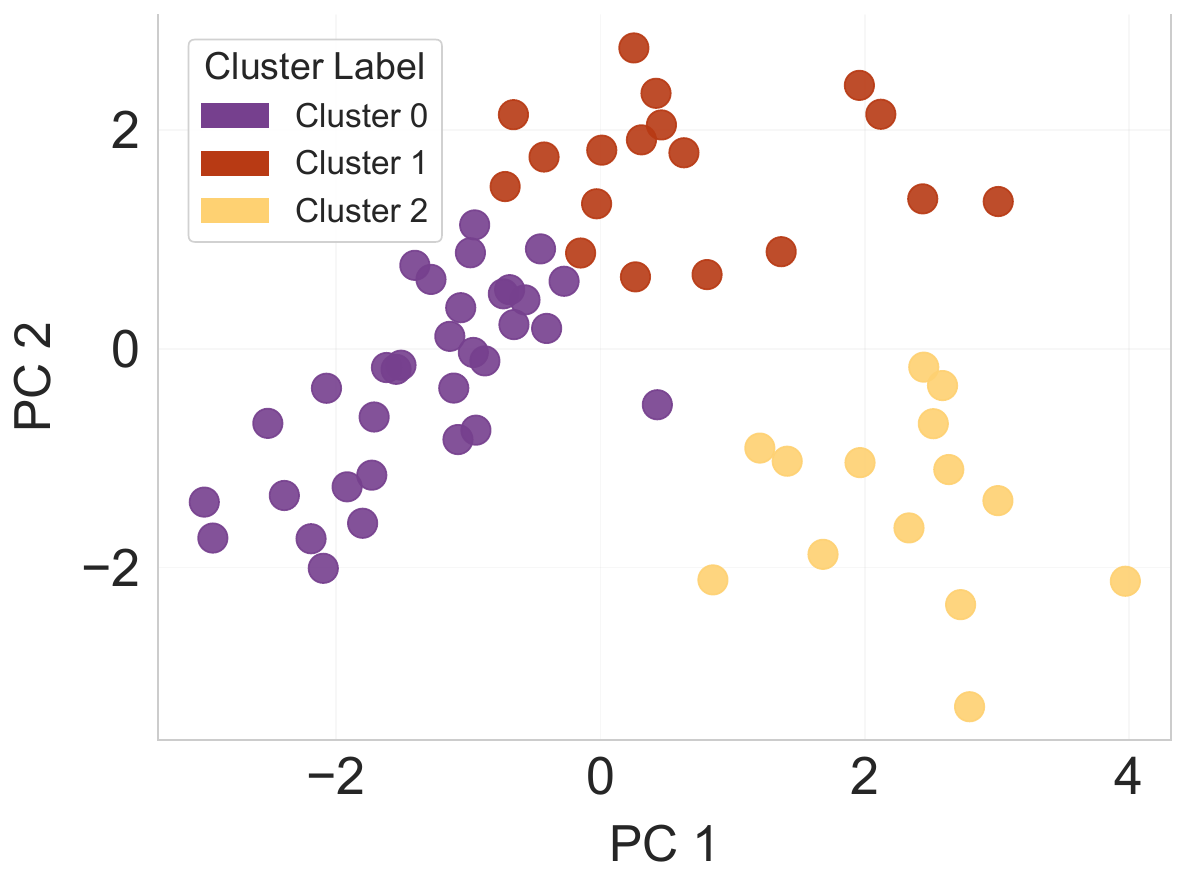}
    \caption{$k$-means clustering of test videos based on comment section features. The projection reveals three distinct clusters, notably Cluster 2 (yellow), which contains videos characterized by high engagement (high PC1) and strong personalization (low PC2).}
    \label{fig:cluster_results}
  \end{minipage}
  \hfill
  \begin{minipage}[t]{0.22\textwidth}
    \includegraphics[width=\textwidth]{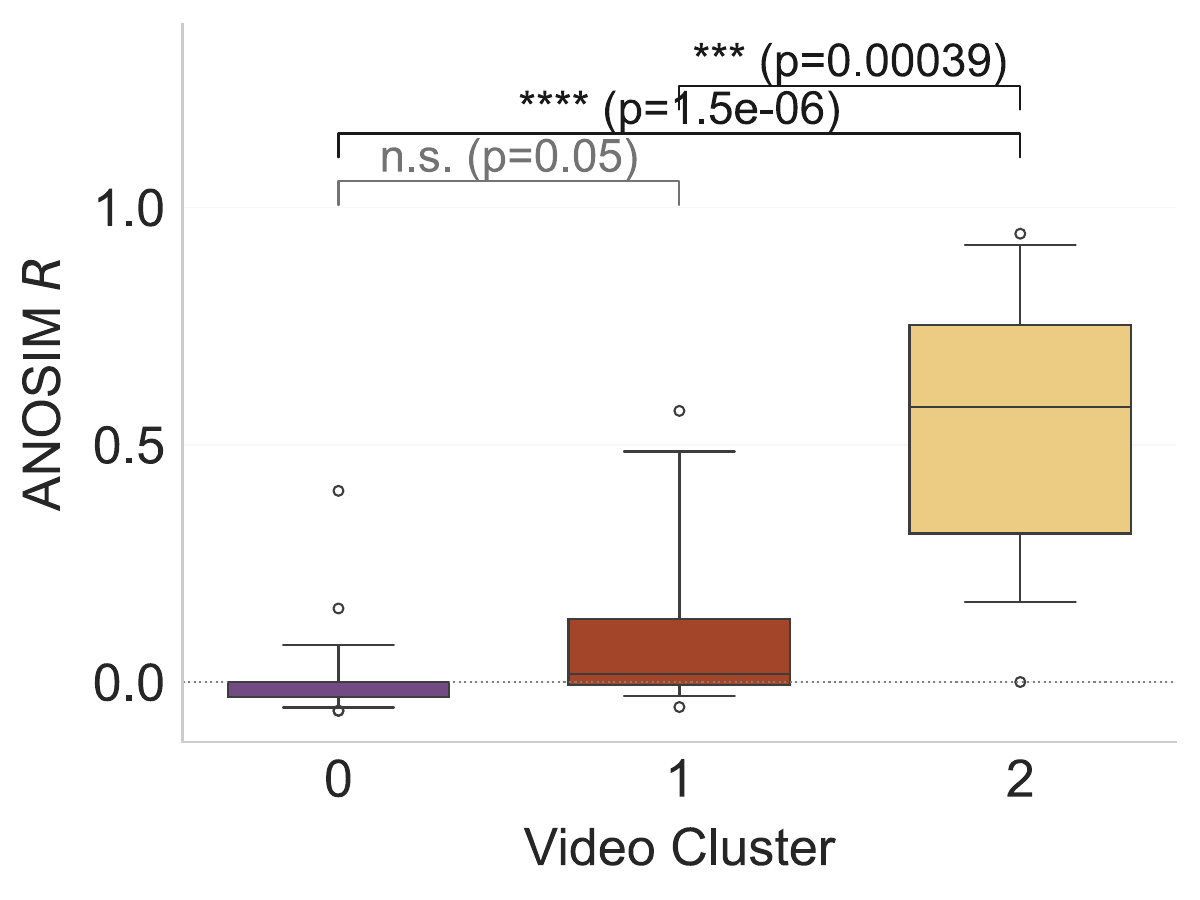}
    \caption{Comparison of $\mathbb{R}_{L,R}$(NDLD) across the three video clusters. Videos in Cluster 2 exhibit significantly higher levels of personalization between $L$ and $R$ accounts compared to Cluster 0 and Cluster 1 (Mann–Whitney U test, $p < 0.001$).}
    \label{fig:boxplot}
  \end{minipage}
\end{figure}

To systematically characterize these variation patterns, we applied Principal Component Analysis (PCA) to the combination of the four comment section features and three between-group $\mathbb{R}$(NDLD) metrics. We excluded $\mathbb{R}$(JD) due to its negligible variation. The resulting five principal components together explain 90\% of the total variance, with the first two components accounting for 68.0\% (41.5\% for PC1 and 26.5\% for PC2). We then used the five PCs as input for $k$-means clustering, which identified three distinct video clusters, as visualized in~\Cref{fig:cluster_results}. \Cref{tab:pca_side} lists the loadings for PC1 and PC2.

\begin{table}[t]
\centering
\begin{tabular}{l r r}
    \toprule
    \textbf{Variable} & \textbf{PC1} & \textbf{PC2} \\
    \midrule
    Volume of discussion          & 0.315 & 0.554 \\
    Inequality of comment likes       & 0.302 & 0.490 \\
    Inequality of comment replies     & 0.250 & 0.447 \\
    Imbalance of partisan discussion   & 0.294 &  -0.180 \\
    $\mathbb{R}_{L,R}$(NDLD) & 0.453 &  -0.290 \\
    $\mathbb{R}_{L,C}$(NDLD) & 0.476 &  -0.272 \\
    $\mathbb{R}_{R,C}$(NDLD) & 0.450 &  -0.252 \\
    \bottomrule
\end{tabular}
  \captionof{table}{Factor loadings for the first two principal components. PC1 captures a general dimension of comment sections, with positive loadings across all variables. PC2 differentiates engagement from personalization, with positive loadings for volume and engagement inequality and negative loadings for the algorithmic ranking metrics $\mathbb{R}$(NDLD).}
  \label{tab:pca_side}
  \end{table}

\begin{itemize}
    \item Cluster 0 (33 videos): Characterized by low PC1 and around-zero PC2. These are low-engagement videos that exhibit little to no comment personalization.
    \item Cluster 1 (18 videos): Characterized by high PC1 and high PC2. These videos exhibit high engagement but limited personalization, likely because the structural engagement features (captured by PC1) and personalization signals (captured by PC2) offset each other.
    \item Cluster 2 (14 videos): Characterized by high PC1 and low PC2. These videos combine high engagement volume with strong personalization effects and imbalanced partisan discussion. 
\end{itemize}

To validate the clustering results, we plot the distribution of $\mathbb{R}_{L,R}$(NDLD) across the three clusters in~\Cref{fig:boxplot}. The analysis shows that videos in Cluster 2 exhibit significantly higher levels of comment personalization between $L$ and $R$ accounts compared to those in Cluster 0 and Cluster 1 (Mann–Whitney U test, $p < 0.001$). Furthermore, for all 14 videos in this cluster, the permutation test $p$-values for $\mathbb{R}$(NDLD) are below 0.01, indicating that the observed differences are statistically significant.

In summary, the results suggest that personalization is structurally driven. When videos exhibit high comment volume and concentrated engagement, both $L$ and $R$ accounts experience rankings that diverge substantially from those of the control group. Moreover, when the videos have an uneven volume of partisan comments, the rankings shown to $L$ and $R$ accounts diverge significantly from each other.

\begin{figure*}[t]
    \centering
    \begin{subfigure}[b]{0.42\textwidth}
        \includegraphics[width=\textwidth]{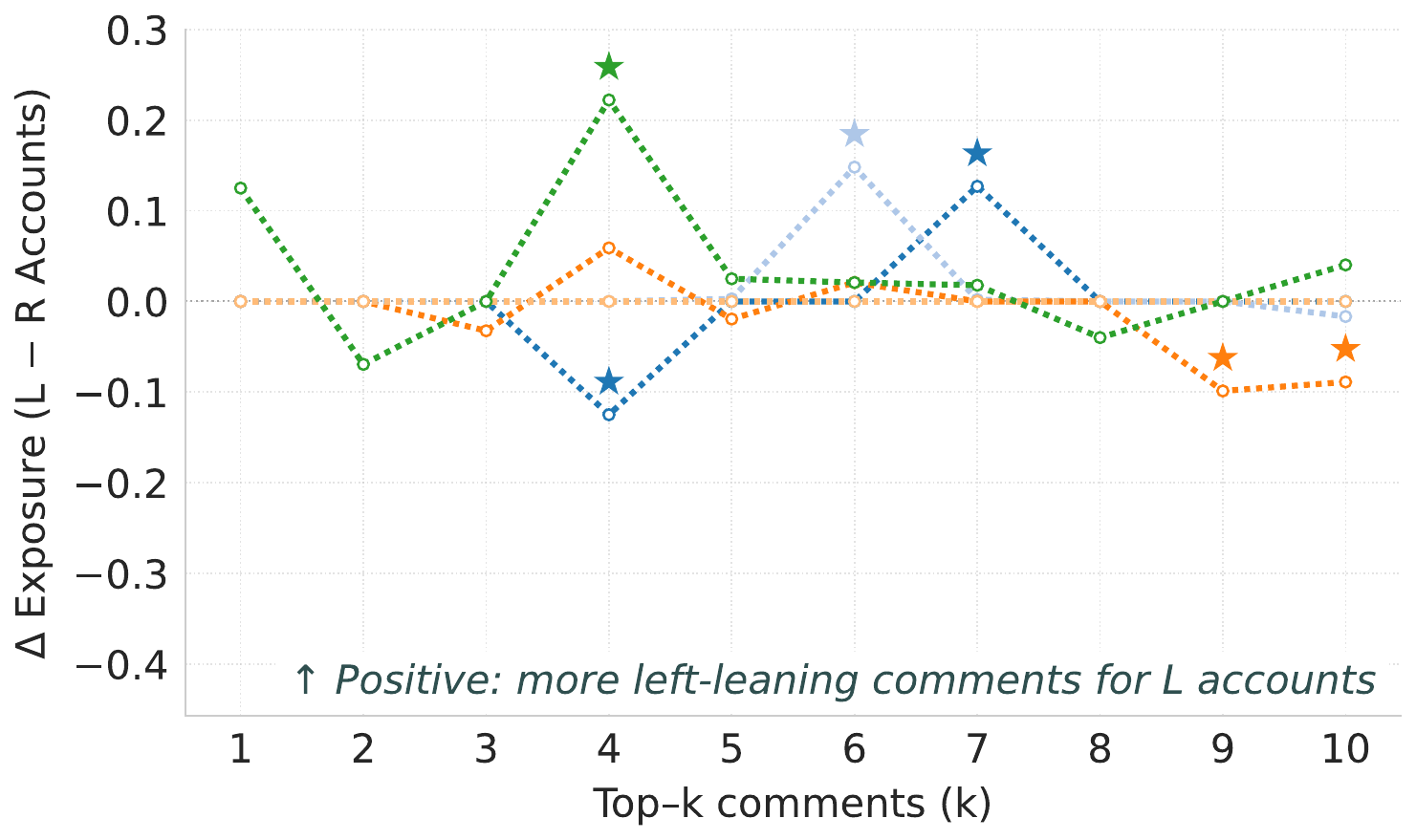}
        \caption{Exposure to left-leaning comments}
        \label{fig:pro_left}
    \end{subfigure}
    \hfill
    \begin{subfigure}[b]{0.42\textwidth}
        \includegraphics[width=\textwidth]{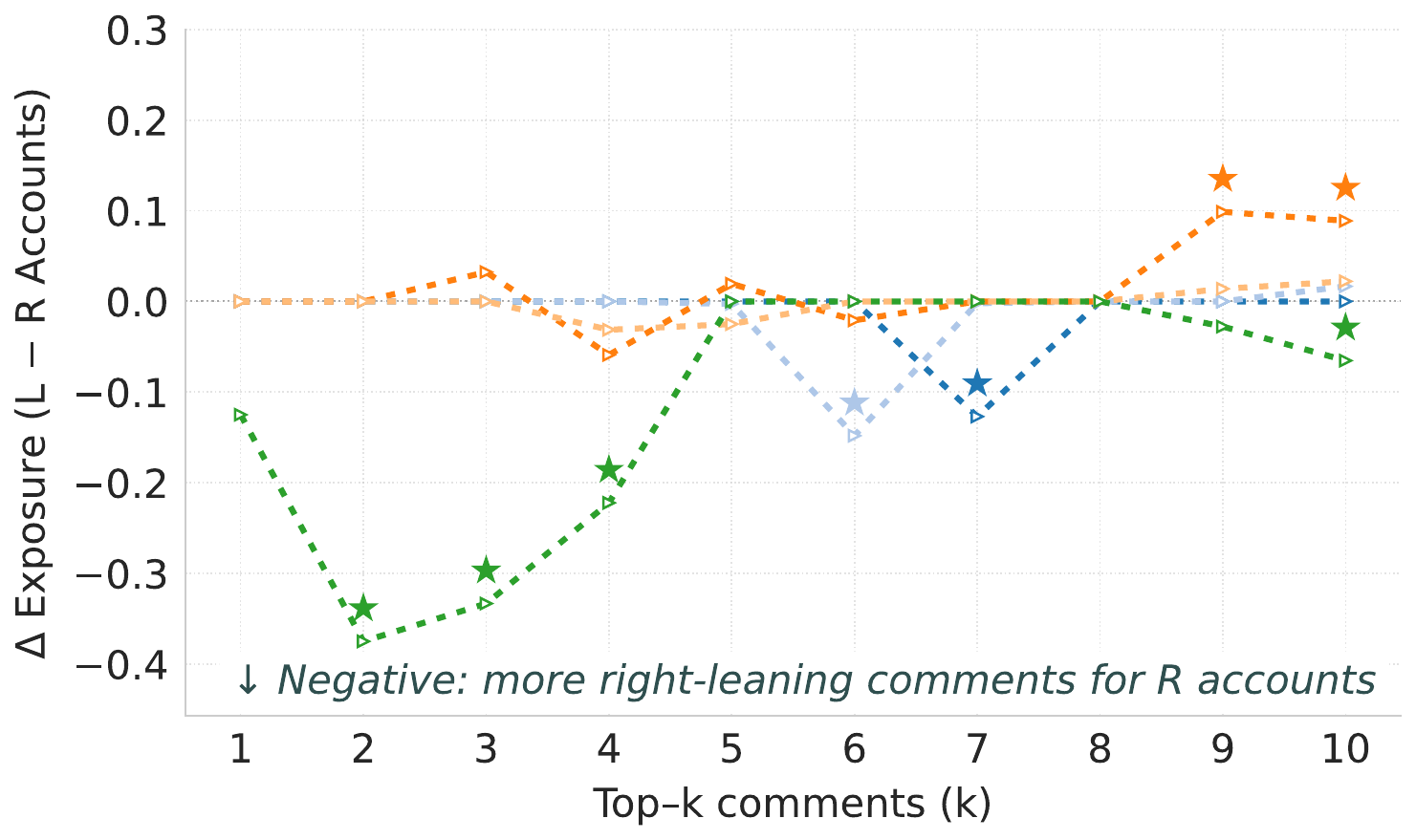}
        \caption{Exposure to right-leaning comments}
        \label{fig:pro_right}
    \end{subfigure}
    \hfill
    \hfill
    \raisebox{1cm}{\scalebox{0.9}{\includegraphics{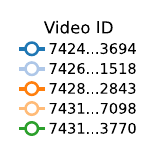}}}
    \caption{The difference of partisan exposure, (a) to left-leaning comments and (b) to right-leaning comments, between nine left-leaning and eight right-leaning accounts at position $k$. Panel (a) exhibits a heterogeneous pattern. On video ``7424...3694'', left-leaning accounts were exposed to more left-leaning comments at position $7$ but more right-leaning comments at position $4$. Panel (b) shows a more consistent pattern of negative deviations at top positions (e.g., ``7431...3770''), indicating that right-leaning accounts are consistently exposed to higher proportions of right-leaning comments than left-leaning accounts.}
\label{fig:lr_comparison}
\end{figure*}

\subsection{RQ2: Does Comment Personalization Favor Assigned Political Leaning?}

To examine whether ranking personalization systematically favors users' existing political identities, we focus on the 14 videos in Cluster 2 that exhibit strong personalization effects in comment ordering.

\subsubsection{Exploratory Case Study Selection} 

To examine the relationship between personalization and political leaning, two authors independently labeled a total of 4,256 comments appearing in the top-$50$ positions across all 22 accounts for these videos. The labeling followed the established codebook (see Appendix~\ref{app:comment_codebook}). Of these comments, 41.9\% were classified as left-leaning, 34.8\% as right-leaning, and 23.3\% as neutral or non-political. We then computed a post-hoc IPD metric, as defined in~\Cref{ssec:video_clustering}, for each video to assess whether the comment pool supported meaningful re-ordering. Specifically, values close to 1 indicate a homogeneous pool dominated by a single political perspective, whereas values close to 0 reflect a more balanced distribution. We retained only videos with a post-hoc IPD below 0.8. This filtering resulted in five videos, all of which originated from the same channel, $@cspanofficial$. 

This reduction in sample size is consistent with the correlation results shown in~\Cref{fig:corr_analysis}, where ranking divergence is positively correlated with the imbalance of partisan discussion. This relationship introduces a selection trade-off: videos exhibiting the strongest algorithmic re-ranking (Cluster 2) tend to be dominated by one-sided political discussions. Consequently, filtering for balanced discussions might exclude a subset of highly personalized videos. Given the small sample size and the distinct comment composition of each video, we analyze the remaining five videos individually rather than aggregating the results. Aggregation would obscure these video-specific patterns. Accordingly, we present this analysis as an exploratory case study.

\subsubsection{Patterns of Differential Partisan Exposure} 

To quantify exposure differences, we computed the proportion of left-leaning and right-leaning comments appearing at the $k$th position ($k = 1$ to $10$) for each partisan account. We then performed a Mann–Whitney U test at each position to assess whether $L$ and $R$ accounts were exposed to significantly different fractions of partisan comments. \Cref{fig:lr_comparison} visualizes the resulting differential exposure. In this plot, positive values indicate greater visibility for $L$ accounts, whereas negative values indicate greater visibility for $R$ accounts. Statistically significant positions are marked with stars. Therefore, if personalization were aligned with users' political leaning, we would expect positive deviations in~\Cref{fig:lr_comparison}(a), where $L$ accounts are exposed to more left-leaning content. Conversely, we would expect negative deviations in~\Cref{fig:lr_comparison}(b), where $L$ accounts are exposed to less right-leaning content (i.e., relatively greater exposure for $R$ accounts).

The results reveal a heterogeneous pattern, with varying degrees of consistency across political groups. In \Cref{fig:lr_comparison}(a), exposure to left-leaning comments exhibits substantial variability. For example, video ``7431...3770'' aligns with the echo chamber hypothesis, showing a significant positive deviation at the fourth position ($k=4$), indicating that $L$ accounts were exposed to more left-leaning comments. In contrast, video ``7424...3694'' shows a significant negative deviation at the same position, indicating that $R$ accounts were exposed to a higher proportion of left-leaning comments than $L$ accounts.

In contrast, \Cref{fig:lr_comparison}(b) shows a more consistent pattern for right-leaning comments. For example, on video ``7431...3770'', significant deviations at top two to four positions are all negative, indicating that right-leaning accounts are systematically prioritized with right-leaning comments. Similarly, videos ``7426...1518'' and ``7424...3694'' exhibit negative deviations in mid-range positions ($k=6$ and $k=7$, respectively). These findings suggest that, while the visibility of left-leaning comments varies across videos, TikTok may more consistently elevate right-leaning perspectives for conservative users on some videos.


\section{Discussion}
\label{sec:discussion}

As the first auditing study to examine personalization effects in comment sections on social platforms, we find that, on TikTok, personalization manifests in the ranking of comments but not in the composition of top comments across different political groups. We also observe that, for certain videos, top-ranked comments are re-ordered in ways that increase users' exposure to politically congenial viewpoints.

For RQ1, we find little evidence of personalization in content composition: the low mean JD ($< 0.08$) across all videos indicates that users, regardless of political leaning, are exposed to a largely shared pool of comments. However, significant personalization emerges in the order of these comments. The ANOSIM analysis shows that between-group differences in comment ranking exceed within-group variation, suggesting that the platform can produce distinct views of the same discussion for different user groups. These findings align with prior auditing studies of search engines, where personalization typically manifests as re-ranking within a stable set of results rather than through the exclusion of content~\cite{robertson2018auditing,krafft2019did,le2019measuring,juneja2023assessing}. Similarly, comment sections are constrained by a finite pool of available comments, leading to compositional stability that contrasts with the higher volatility observed in feed recommendations. A robustness check further shows that this effect diminishes with depth, consistent with prior work finding that algorithmic influence is concentrated in top-ranked, high-attention positions~\cite{robertson2018auditing}. Taken together, these results suggest that while the underlying discussion in TikTok comment sections remains intact, users' ``perceived reality'' can be shaped through comment re-ranking. By altering the order of top comments, the platform creates distinct first impressions for different user groups, even when they are watching the same video.

We further investigate the structural conditions that drive algorithmic re-ranking by analyzing features such as comment volume, engagement inequality, and content diversity. We find that videos with larger comment volumes, highly concentrated engagement, and strongly imbalanced partisan discussions tend to exhibit more pronounced personalization effects. The correlation with comment volume is intuitive, as a larger pool of comments provides greater flexibility for re-ranking. More notably, the association with engagement inequality suggests a potential algorithmic counterbalance to popularity bias. One possible interpretation is that when a small number of popular comments dominate user attention, the system diversifies exposure by surfacing personalized alternatives that would otherwise be suppressed by ``rich-get-richer'' dynamics in like counts~\cite{abdollahpouri2019managing}. Most surprisingly, we observe a strong relationship between imbalance of partisan discussion and ranking divergence. This pattern suggests that the algorithm more aggressively reorders comments in one-sided discussions to produce divergent viewing experiences, potentially as a mechanism to simulate a more balanced discourse.

Beyond overall differences in comment ranking, we examine whether this re-ranking systematically increases each audit group's exposure to politically congenial content. However, our findings reveal a structural paradox that complicates this assessment: the conditions that trigger the strongest algorithmic intervention (high partisan imbalance in the comment sections) also limit the potential for differentiated partisan exposure. When discussions are highly homogeneous, the algorithm may re-rank more aggressively, but the lack of alternative viewpoints constrains the extent to which divergent exposures can emerge. Consequently, we shift from a quantitative analysis to an exploratory case study of five test videos that exhibit both strong personalization potential and sufficient cross-partisan discussion.

Our results for RQ2 suggest that TikTok's comment ranking can, in specific contexts, function as an exposure-shaping mechanism in political discussions, particularly by elevating the visibility of congenial viewpoints for right-leaning users. This asymmetry likely stems from an underlying imbalance in the comment pool. As our analysis of center videos indicates a prevailing left-leaning skew, the algorithm requires little intervention to expose users to left-leaning content, which is already abundant. In contrast, to ensure that right-leaning users encounter congenial perspectives, the system must actively re-rank and elevate these relatively scarce viewpoints. This dynamic helps explain why the signal for right-leaning personalization, as shown in~\Cref{fig:lr_comparison}(b), is more consistent, whereas left-leaning exposure patterns remain more variable. This asymmetry is particularly concerning given the hierarchical nature of attention in comment sections, where top-ranked comments disproportionately shape users' perceptions of dominant viewpoints and social consensus~\cite{Lee2012CommentsPublicOpinion,northcutt2017comment}. At the same time, such partisan selective visibility is not uniform across all content. Our analysis is based on a small set of five videos, and thus should be interpreted as exploratory. Nevertheless, the findings align with prior literature showing that personalization is highly context-dependent, varying across factors such as search queries~\cite{le2019measuring}, content domains~\cite{tomlein2021audit,gong2025clipmind}, and communities~\cite{mok2023echo}. For example, \citet{le2019measuring} found that only nine out of 50 queries exhibited measurable personalization.

Taken together with the findings from RQ1, our results suggest that TikTok manifests a subtle form of algorithmic re-ranking rather than outright content exclusion across political groups. While the overall composition of comment sections remains consistent across accounts, indicating that the platform does not enforce a ``hard'' filter bubble by removing opposing viewpoints, the algorithm might selectively elevate congenial partisan comments into highly visible top positions for a subset of videos. Even in the absence of explicit censorship, this ranking mechanism can reinforce polarization by shaping the salience of information, such that users' initial impressions of public discourse align with their existing beliefs. This process gives rise to a localized ``soft'' echo chamber, which may prime users' perceptions of public sentiment before they engage with the broader and more heterogeneous comment thread.

Practically, how can comment personalization be implemented given the challenges of modeling large-scale, rapidly evolving comment corpora? \citet{piccardi2025reranking} propose an LLM-aided browser extension that enables real-time re-ranking of social media feeds. Building on this idea, platforms or even third-party developers could leverage LLMs to dynamically re-rank content in the comment sections, for example, by up-ranking or down-ranking politically congenial comments. Alternatively, this problem can be framed as a classical recommendation task~\cite{hsu2009ranking}, where user and comment embeddings are learned in advance and comments are prioritized based on their predicted likelihood of user engagement, where engagement can be defined as leaving a reply or giving a like.

\paragraph{Ethical considerations}
For the auditing experiment, we collected only observational data from TikTok using sock-puppet accounts and did not involve any real users. All analyzed videos were publicly available on the platform. The accounts were limited to performing search and watch actions and did not engage in other forms of interactions (e.g., posting, liking, or commenting) that could affect other users. As a result, the study did not influence the experience of real users. All sock-puppet accounts were deleted after the experiment, further minimizing the risk of misuse.

\paragraph{Limitations and future work}
First, our training strategy relies exclusively on ``search and watch'' behaviors to evaluate algorithmic personalization. While we deliberately exclude explicit signals like ``follows'' and ``likes'' to minimize potential impacts on public engagement metrics, prior work suggests that factors such as location, follows, likes, and direct interactions with comments also contribute to personalization. Consequently, our findings likely represent a conservative estimate of the extent of algorithmic tailoring. Future research should systematically isolate the effects of these active engagement signals using alternative audit designs, particularly those capable of evaluating ``follow'' dynamics without permanently altering public platform data.

Second, despite our efforts to select balanced videos, our post-hoc analysis reveals a persistent left-leaning skew in baseline comment composition. This imbalance likely obscures the true extent of personalization for left-leaning accounts. Since the default ranking already surfaces congenial content for these users, the algorithm requires minimal re-ranking to satisfy their preferences, potentially leading to an underestimate of personalization effects. Future studies should employ a broader sampling strategy that spans a wider range of cross-partisan discussion. In addition, examining temporal dynamics is crucial. Because our audit focuses on older videos with relatively stable comment sections, it may not capture the higher volatility and rapid re-ranking characteristics of live or breaking news contexts.

Third, our study examines a limited sample of 65 videos across 22 accounts, with the exploratory analysis of political alignment further restricted to five videos. While these case studies provide preliminary evidence of echo chamber effects in top positions, they do not allow us to assess the prevalence of this phenomenon across the broader platform. Our scale is constrained by the resource-intensive nature of manual annotation and sock-puppet management. Future work should leverage automated, large-scale auditing frameworks to evaluate whether these re-ranking patterns are systematic features of the platform's political ecosystem or are confined to specific high-engagement contexts.


\section{Conclusion}
\label{sec:conclusion}

We contribute the first auditing study of personalization effects in TikTok comment sections. Using 22 sock-puppet accounts to interact with videos related to the 2024 U.S. presidential election, we show that personalization manifests primarily through the ranking of comments, particularly at top positions. Specifically, videos with high comment volume, unequal engagement distribution, and strong imbalance of partisan discussion exhibit the most pronounced algorithmic re-ranking. Regarding political alignment, our exploratory analysis reveals that this mechanism can shape the visibility of certain political viewpoints. Overall, this work establishes a foundation for systematic auditing of personalization effects in comment sections on social platforms.
 


\bibliography{references}

\appendix

\section{Annotation Codebook}
\label{app:comment_codebook}

\begin{table}[H]
\small
  \centering
  \begin{tabular}{r p{4.7cm}}
    \toprule
    \textbf{Label} & \textbf{Definition and examples} \\
    \midrule
    \textbf{Pro-Democrat} & Endorsing political figures (e.g., Kamala Harris, Tim Walz, Barack Obama) or values (e.g., Black Lives Matter, abortion and immigration rights) associated with the Democratic Party.\\ 
    \textbf{Anti-Democrat} & Criticizing political figures or values associated with the Democratic Party. \\
    \textbf{Pro-Republican} & Endorsing political figures (e.g., Donald Trump, JD Vance, Ted Cruz) or values (e.g., Christianity, gun rights) associated with the Republican Party.\\ 
    \textbf{Anti-Republican} & Criticizing political figures or values associated with the Republican Party. \\
    \textbf{Neutral} & Expressing unclear stance or non-political content. \\
    \bottomrule
  \end{tabular}
  \label{tab:comment_codebook}
  \caption{Codebook for labeling comment political stance}
\end{table}

\section{LLM Prompt}
\label{app:prompt}

The following prompt is used to classify political stance of TikTok comments via GPT-4o. The input includes both the comment text and the corresponding video description.  

\begin{lstlisting}[style=promptstyle][escapeinside=||]
You are an expert in political discourse analysis. Your task is to boldly and decisively classify the political leaning of TikTok comments. 
Use the video description for context, and analyze linguistic cues, pronouns, and especially emojis, which should be treated as strong indicators of sentiment and political alignment. 
For example, |\textcolor{blue}{$\heartsuit$}| for pro-democrat and
|\textcolor{red}{$\heartsuit$}| for pro-republican.
Frame your analysis within the context of the 2024 U.S. presidential election. Assume that comments are likely to carry political intent-even subtle cues can signal alignment.
Use the following criteria for classification:
Pro-Democrat: endorse political figures from Democrat party, e.g., Kamala Harris, Tim Walz, Barack Obama, Nancy Pelosi;
Anti-Democrat: attack political figures from Democrat party;
Pro-Republican: endorse political figures from Republican party, e.g., Trump, Vance, Ted Cruz;
Anti-Republican: attack political figures from Republican party;
Neutral: non-political, or targeting people with no clear political affiliations, or do not express political stances.
A comment can be both Pro-Republican and Anti-Democrat, in such cases, prioritize Pro-Republican over Anti-Democrat. For example, if a comment endorses Trump and defames Harris, choose Pro-Republican.After examining these factors, make a bold and decisive classification—do not hedge your answer. Respond with one of the following labels only:
Pro-Democrat, Anti-Democrat, Pro-Republican, Anti-Republican, or Neutral.
Few-shot examples:
---
**Video Description:** 'Donald Trump spends his time attacking great American cities full of hardworking people. He’s not looking out for you.'
**Comment:** 'cant wait for madam president'
**Classification:** Pro-Democrat

Now, classify the following comment:
**Video Description:** {row['video_description']}
**Comment:** {row['comment']}
**Classification:**"""

\end{lstlisting}

\end{document}